\documentstyle{article}
\font\tenbf=cmbx10
\font\tenrm=cmr10
\font\tenit=cmti10
\font\elevenbf=cmbx10 scaled\magstep 1
\font\elevenrm=cmr10 scaled\magstep 1
\font\elevenit=cmti10 scaled\magstep 1

\font\ninerm=cmr9

\begin{document}
\begin{center}{\tenbf EFFECTIVE FIELD THEORY OF MULTIPARTICLE
CORRELATIONS\footnote{\ninerm\baselineskip=11pt
Talk presented by I. Sarcevic at the XXIII
International Symposium on Multiparticle
Dynamics, Aspen, CO, September 12-17, 1993.}}

               \vglue 10pt

\vglue 5pt
\vglue 1.0cm
\vglue 0.3cm
\centerline{\tenrm Ina Sarcevic }
\baselineskip=13pt
\centerline {\tenit Department of Physics, University of Arizona}
\baselineskip=12pt
\centerline{\tenit Tucson, AZ 85721}
\vglue 0.3cm
\centerline {\tenrm H.C.\ Eggers}
\centerline {\tenit Department of Physics, McGill University}
\centerline{\tenit Montreal, Quebec, Canada H3A 2T8}
\vglue 0.3cm
\centerline {\tenrm H.\ Th.\ Elze }
\centerline {\tenit CERN-TH, CH-1211 Geneva 23, Switzerland}
\end{center}
\vglue 0.8cm
\begin{center}
{\tenrm ABSTRACT}
\end{center}
\vglue 0.3cm
{\rightskip=3pc
 \leftskip=3pc
 \tenrm\baselineskip=12pt
 \noindent
We present an effective field theory of
multiparticle correlations based on analogy with Ginzburg-Landau
theory of superconductivity.  We assume that the field
represents
particle density fluctuations, and show that
in the case of
gaussian-type
effective action
there are
no higher-order
correlations, in agreement with the recent data.  We predict that
two-particle correlations have Yukawa form.  We also present our
results for the
two-dimensional and one-dimensional
two-particle correlations (i.e. cumulants)
as projections of our theory to lower dimensions.
\vglue 0.6cm}
{\elevenbf\noindent 1. Introduction}
\vglue 0.2cm
\baselineskip=14pt
\elevenrm
Recent experimental measurements of fluctuations in
transverse energy
in ultrarelativistic heavy-ion collisions
have provided evidence that nuclear constituents scatter and
produce particles coherently [1].
The observed
fluctuations in the deposited energy and multiplicity are
remarkably large and can not be described by an
independent-collision model [2].
The same conclusion has been reached when
multiparticle density
fluctuations in different
phase space regions have been studied
via
factorial moments, defined as
$$F_{q}(\delta y) = {1\over M} \sum_{m=1}^{M} {\langle n_{m}(n_{m}-1)
\ldots (n_{m}-q+1)\rangle \over  \langle n_{m} \rangle^{q}},\eqno(1)$$
where M is
the number of rapidity bins $(\delta y = Y/M )$ and $n_m$ is the
number of particles in the $m^{th}$ bin [3].
In the past few years,
these moments have
been measured for different targets and
projectiles
at energy of
$200$GeV/nucleon
[4].  They
were found to
increase
with decreasing bin size
indicating nonstatistical fluctuations and being
incompatible with
the predictions of the
standard
classical hadronization models embedded in the existing Monte Carlo
models [4].  In addition,
the observed effect, sometimes referred to as the
``intermittency effect'', can not be accounted for by the
superposition of independent nucleon-nucleon collisions, even
when rescattering and geometrical effects are included [3].

The possibility of creating
the new form of matter, the quark-gluon plasma, in high-energy
heavy-ion collisions
have inspired intensive
theoretical work on identifying
the unambiguous QGP signal.
The
observation of the unusually large
multiparticle density
fluctuations have created a new excitement in the field,
especially as a possibility of
pointing towards the
onset of the
phase transition from quark-gluon plasma to hadronic matter.
Phase transitions in QCD at high
temperatures are of general interest -- they are
directly relevant to cosmology, since such a phase transition
occurred throughout the universe during the
early moments of the big bang and
a first order phase
transition could have
altered primordial nuclear abundances.
Unfortunately,
up to now
there
are no conclusive predictions for detecting the
quark
matter in heavy-ion collisions
and there is no
theory
to describe the observed ``intermittency'' phenomenon [5].
\vglue 0.6cm
{\elevenbf\noindent 2. Multiparticle Correlations in Heavy-Ion Collisions
}
\vglue 0.4cm

Multiparticle correlations in
three ``dimensions'' are usually measured in the following way.
A given total interval
$\Omega_{\rm tot} = \Delta Y\, \Delta \phi\, \Delta P$ is
subdivided into $M^3$ bins of side lengths
$(\Delta Y/M,\, \Delta \phi/M,\, \Delta P/M)$. With $n_{klm}$
the number of particles in bin $(k,l,m)$
and $n^{[q]} = n!/(n-q)!$
the ``vertical'' factorial moment is
$$F_q^v(M) \equiv {1\over M^3}
\sum_{k,l,m=1}^M
{
\langle n_{klm}^{[q]} \rangle
\over
\langle n_{klm} \rangle^{q}
}
=
{1\over M^3}
\sum_{k,l,m=1}^M
{
\int_{\Omega_{klm}} \prod_i  d^3 x_i \, \rho_q(x_1\ldots x_q)
\over
\left[\int_{\Omega_{klm}} d^3 x \, \rho_1(x)\right]^q
}. \eqno(2)$$
\noindent
The second equality illustrates how the factorial moment can be
written in terms of integrals of the correlation function [6].
The alternative ``horizontal'' factorial moment is often
preferred for three-dimensional analysis; this form,
while being much more stable, has the
drawback that it depends on the shape of the one-particle
distribution function $\rho_1$.

To measure true particle correlations, known as cumulants, the
trivial background must be subtracted. The first two cumulants
are
$$C_2(x_1,x_2) = \rho_2(x_1,x_2) - \rho_1(x_1) \rho_1(x_2)\eqno(3)$$
$$C_3(x_1,x_2,x_3) = \rho_3(x_1,x_2,x_3) - \sum_{perm}
\rho_1(x_1)\rho_2(x_2,x_3)
+ 2 \rho_1(x_1)\rho_1(x_2)\rho_1(x_3).$$
By integrating these relations over each bin, one can
derive equations for integrated cumulants
$K_q^v = \int \prod dx_k C_q / (\int dx \rho_1(x))^q$
in terms of the above vertical factorial moments [7],
$$K_2^v = F_2^v - 1, \ \ \ \ \ \ K_3^v = F_3^v - 3F_2^v + 2
\ \ \ \ \mbox{etc}.
\eqno(4)$$
Whenever there are no true correlations, these cumulants become zero.
It was found that in the case of heavy-ion
collisions, there are only two-particle correlations: while $K_2$
is positive, the values of $K_3$, $K_4$ and $K_5$
have been found to be consistent with zero [8].
First found in terms
of one-dimensional rapidity data, this has been confirmed by
measurements by NA35 in two and three dimensions [9].
Corresponding findings for other nuclei and energies were
published before [10].
\vglue 0.4cm

\vglue 0.4cm
{\elevenbf \noindent 3. Effective Field Theory of
Multiparticle Correlations
}
\vglue 0.5cm

We have seen that particles
produced in high-energy heavy-ion collisions exhibit
only two-particle correlations, indicating that perhaps
higher-order correlations are washed out by rescattering
of the initially correlated particles.  Presently, there is
 no theory that describes this phenomena.
Recently, we have proposed
a three-dimensional
statistical field theory of density
fluctuations which has these features [11,12].
This model was formulated in analogy with the Ginzburg-Landau
theory of superconductivity.  The large
number of particles produced in ultrarelativistic
heavy-ion collisions justifies the use if a
statistical theory of particle production.
The formal analogy with the statistical mechanics
of a one-dimensional ``gas'' was first pointed out by
Feynman and Wilson and was
later further developed by
Scalapino and Sugar [13] and many others [5].
The idea is to build a
statistical theory of the macroscopic observables by
imagining that the microscopic degrees of freedom are
integrated out and represented in terms of a few
phenomenological parameters
and by postulating that
this theory
will
eventually be derived from a more
fundamental theory, such
as QCD.
\par
While in the G-L theory of superconductivity
the field (i.e. the order parameter) represents superconducting
pairs, in the particle production problem, the relevant
variable is the density fluctuation.
We define a random field $\Phi$ as a function in a
three-dimensional space spanned by $(y,\phi,p_\perp)$.
Throughout, $p_\perp$ will be implicitly divided by a constant
scale $\cal P$ so that it is dimensionless. Since we are not
looking for a phase transition, we omit the quartic term
and start with the functional [11]
$$
F[\Phi] = \int_0^P dy\, \int_{-P/2}^{P/2} d^2p_\perp
\left[ a^2 \left(\partial\Phi / \partial y\right)^2
     + a^2 \left(\nabla_{\vec p_\perp} \Phi \right)^2
     + \mu^2 \Phi^2
\right] \;. \eqno(5)$$
Taking the appropriate functional derivative, we find for the functional
(5) the three-dimensional form of the two-point function
$$
\langle \Phi(\vec x_1)\Phi(\vec x_2) \rangle
= {1\over 8 \pi a^2} {e^{-R/\xi} \over R} \;,
\eqno(6)$$
where $\xi = a/\mu$ and
$
R  \equiv
[(y_1-y_2)^2 + p_{\perp 1}^2 + p_{\perp 2}^2
- 2p_{\perp 1} p_{\perp 2}\cos(\phi_1-\phi_2) ]^{1/2}
$.
Further, we
define $\Phi(\vec x)$ as the fluctuation at the point $\vec x$
of the particle density for a given event, $\hat\rho_1(\vec x)$,
above/below the mean single particle distribution $\rho_1$ at that
point:
$$
\Phi(\vec x) \equiv {\hat\rho_1(\vec x) \over \rho_1(\vec x) } - 1 \;.
\eqno(7)$$
Through these definitions, we find that
$
\langle \Phi(\vec x_1)\Phi(\vec x_2) \rangle \;
= \;  k_2(\vec x_1,\vec x_2)
$
and that all higher order cumulants become exactly
zero,  $k_{q\ge 3} = 0$.
By means of the specific form of the
functional (5) and the definition of $\Phi$ as a
fluctuation, we take account of the experimental facts in this
regard. What is not experimentally certain and is to be tested is
whether the second order correlations obey the Yukawa form
(6).

\vglue 0.5cm
{\elevenbf \noindent 4. Projections of
Multiparticle Correlations (i.e. Cumulants) to Lower Dimensions}
\vglue 0.4cm
The second reduced cumulant $k_2 \propto e^{-R/\xi}/R$ can
be compared to data only after a suitable integration over
its variables. For three dimensions, the vertical
integrated cumulant is given by
$K_2^v(\delta y,\delta\phi,\delta p)
=  F_2^v - 1
=  M^{-3}\sum_{k,l,m=1}^M K_2^v(k,l,m) \;,
$
(always taking $\vec x \equiv (y,\phi,p_\perp)$), with
$$
K_2^v(k,l,m)    =
{
\int_{\Omega_{klm}}d^3\vec x_1\, d^3 \vec x_2\; C_2(\vec x_1,\vec x_2)
\over
\left[
\int_{\Omega_{klm}}d^3 \vec x\; \rho_1(\vec x)
\right]^2
}
=
\int_{\Omega_{klm}} d^3\vec x_1\, d^3\vec x_2
{
k_2(\vec x_1,\vec x_2) \rho_1(\vec x_1) \rho_1(\vec x_2)
\over
\left[\int_{\Omega_{klm}} d^3 \vec x\, \rho_1(\vec x)
\right]^2
} \;,
\eqno(8)$$
i.e.\ the integration of $k_2$ involves a correction due to the shape of
the one-particle three-dimensional distribution function
$\rho_1(\vec x)$. Eq.\ (8) as it stands is exact;
horizontal versions have also been derived [12].
A first test
of our model would therefore be to see if Eq. (4) or its
horizontal equivalent obeys the data in $(y,\phi,p_\perp)$.

The theoretical $k_2(\vec x_1,\vec x_2)$ is further tested by
comparing to factorial cumulant data of lower dimensions.
For example, in $(y,\phi)$, the cumulant is
$K_2^v(\delta y,\delta\phi) = M^{-2} \sum_{lm} K_2^v(l,m)$ with
$p_\perp$ integrated over the whole window $\Delta P$,
$$
K_2^v(l,m) =
\int_{\Omega_m}dy_1 dy_2 \int_{\Omega_l}d\phi_1 d\phi_2
\int_{\Delta P} dp_1 dp_2\,
{
 k_2(\vec x_1,\vec x_2)\, \rho_1(\vec x_1)\rho_1(\vec x_2)
\over
\left[ \int_{\Omega_m}dy \int_{\Omega_l}d\phi \int_{\Delta P} dp\;
\rho_1(\vec x) \right]^2
} .
\eqno(9)$$
Cumulants of other variable combinations and lower dimensions
are obtained analogously.
With these relations it is thus possible, given any
three-dimensional theoretical function $k_2$ (or $r_2$), to
compute factorial cumulants and moments for any combination of
its variables. Doing this for different variables serves as
a strong test of the theoretical function as the moments probe
its different regions.

In Figures  2-3  (see Ref. 12)
we present our results for the vertical and
horizontal factorial moments.
In our calculation of the projections, we have made the
following approximations:
We factorize the one-particle distribution into its separate variables:
$\rho_1(\vec x) = \langle N \rangle_\Omega\,
                  g(y)\, h(\phi)\, f(p_\perp)$,
where the three distributions $g$, $h$ and $f$ are separately normalized
over their respective total intervals $\Delta Y$, $\Delta \Phi$
and $\Delta P$.
The azimuthal distribution is taken as flat, $h(\phi) = 1/\Delta\Phi$.
We use the full experimental parametrization for $f(p_\perp)$
provided by NA35.
The choice of
two parameters $a$ and $\xi$ in Figures 2-3 [12]
are given
only as an illustration.
They should be determined by comparison with
three-dimensional data.
Once they are fixed, the two-dimensional and one-dimensional
projections are genuine predictions of the theory.

In summary, we have presented a three-dimensional
effective
field theory of multiparticle correlations, which gives
no higer-order
correlations, in agreement with the recent heavy-ion data.
In our theory, two-particle
correlations have Yukawa form and the corresponding integrated
cumulants have singular behavior for small regions of phase space.
This prediction seems to be in qualitative agreement with the
recent
NA35 data [14].
In addition, we have shown that once the parameters $a$ and $\xi$ are
determined from comparison with three-dimensional data,
our theory gives genuine predictions
for the two-dimensional and one-dimensional cumulants.
It will be interesting to see whether all
our predictions are
confirmed with recent NA35 data
presented
at this conference [14].

\vglue 0.5cm
{\elevenbf \noindent 5. Acknowledgements \hfil}
\vglue 0.4cm
This work was supported in part by the
DOE grants DE-FG03-93ER40792 and
DE-FG02-85ER40213.

\vglue 0.5cm
{\elevenbf\noindent 6. References \hfil}
\vglue 0.4cm
\begin{enumerate}

\item
NA34 Collaboration, F. Corriveau {\elevenit et al.},
{\elevenit Z. Phys.} {\elevenbf C38}, 15 (1988); NA34 Collaboration,
J. Schukraft {\elevenit et al.} {\elevenit Z. Phys.} {\elevenbf C38},
59 (1988).

\item
G. Baym, G. Friedman and I. Sarcevic, {\elevenit
Phys. Lett.} {\elevenbf B219}, 205 (1989).

\item
A. Capella, K. Fialkowski and A. Krzywicki,
{\elevenit Phys. Lett.} {\elevenbf B230}, 149 (1989).

\item
For a recent experimental review, see W. Kittel, these
Proceedings.

\item
For a recent theoretical review, see R. Hwa, these Proceedings.

\item
P. Carruthers and I. Sarcevic, {\elevenit Phys. Rev. Lett.}
{\elevenbf 63}, 1562 (1989).
\item
P.\ Carruthers, H.C.\ Eggers, and I.\ Sarcevic,
{\elevenit Phys. Lett.} {\elevenbf 254B}, 258 (1991).

\item
P.\ Carruthers, H.C.\ Eggers
and I.Sarcevic,
{\elevenit  Phys.\ Rev.} {\elevenbf C44}, 1629 (1991).

\item
NA35 Collaboration, I.\ Derado, in
{\elevenit Proceedings of the
Ringberg Workshop on
Multiparticle Production},
ed.\ R.C.\ Hwa, W.\ Ochs and N.\ Schmitz
(World Scientific, 1992) pg. 184;
NA35 Collaboration, P.\ Seyboth {\elevenit et al.}
, {\elevenit Nucl. Phys.} {\elevenbf A544}
293 (1992).

\item
H. C. Eggers, Ph. D. Thesis, University of Arizona, 1991.

\item
H.-Th.\ Elze and I.\ Sarcevic,
{\elevenit Phys. Rev. Lett.}  {\elevenbf 68}, 1988 (1992).

\item
H.C.\ Eggers, H.T.\ Elze and I.\ Sarcevic,
preprint TPR-92-41 (to be published).

\item
D. J. Scalapino and R.\ L.\ Sugar,
               {\elevenit Phys.\ Rev.} {\elevenbf D8}, 2284 (1973);
             J.C.\ Botke, D.J.
Scalapino and R.L.\ Sugar,
{\elevenit ibid.}, {\elevenbf D9}, 813 (1974);
{\elevenit ibid.}, {\elevenbf 10}, 1604 (1974).
\item
NA35 Collaboration, P. Seyboth, these Proceedings.

\end{enumerate}

\end{document}